# Investigation of the effect of $Au_2O_3$ dopant on elastic properties of $PbO$-$B_2O_3$-$SeO_2$: $Er_2O_3$ glass ceramics by ultrasonic techniques

A.Siva Sesha Reddy [a], A.V.Kityk [b], J.Jedryk [b,*], P.Rakus [b], A.Wojciechowski [b], N.Venkatramaiah [c], V.Ravi Kumar [a], N.Veeraiah [a,*]

[a] Department of Physics, Acharya Nagarjuna University, Nagarjuna Nagar, Andhra Pradesh 522 510, India
[b] Faculty of Electrical Engineering, Czestochowa University of Technology, Armii Krajowej 17, Czestochowa PL-42-201, Poland
[c] Department of Chemistry, SRM Institute of Science and Technology, Kattankulathur, Chennai, Tamil Nadu 603203, India

ABSTRACT

Various elastic coefficients of $Au_2O_3$ doped $PbO$-$B_2O_3$-$SeO_2$:$Er_2O_3$ (PBSE) glass ceramics were evaluated as functions of $Au_2O_3$ content using ultrasonic velocity measurements. The elastic coefficients and micro-hardness showed a decreasing tendency with the concentration of $Au_2O_3$. Such decrease is attributed to the increasing concentration of gold metallic particles and $[SeO_3]^{2-}$ groups that acted as modifiers and induced imperfections in these samples. Obtained results were observed to be consistent with the conclusions drawn from spectroscopic studies that include X-ray photoelectron spectroscopy (XPS), infrared (IR), photoluminescence (PL) and positron annihilation (PAL) spectroscopy studies. Overall, these studies have revealed that even though, the presence of gold metallic particles is preferable for achieving superior luminescence and electrical properties, presence of such particles caused to decrease the elastic coefficients and micro-hardness of these glass ceramics. However, when the concentration of $Au_2O_3$ is increased beyond 0.075 mol%, we have observed a slight increase of elastic coefficients and micro-hardness.

## 1. Introduction

Various elastic coefficients like Young's modulus ($Y$), shear modulus ($G$), Poisson's ratio ($\sigma$), micro-hardness ($H$) and various other parameters like Debye temperature ($\theta_D$) etc., can be evaluated by determining sound velocities in glasses and glass ceramics. Knowledge of such parameters not only helps in assessing their elastic behavior but also throw useful information on structural variations occurring in the glasses and glass ceramics. In the recent past, a large number of studies on elastic properties of different glasses and glass ceramics were investigated by measuring ultrasonic velocities in them [1-5].

Among various glass systems, selenium oxide ($SeO_2$) is a rare glass system. As such it is a feeble glass forming oxide and forms the glass when the modifying oxides like PbO and different traditional glass formers like borate, silicate, etc. are added. Interestingly, in the process of glass forming, the coordination number of Se ions with oxygen varies from 4 to 3 and the oxide of selenium transforms from $[SeO_4]^{2-}$ (selenate) structural units to isolated $[SeO_3]^{2-}$ (selenite) pyramidal units [6, 7]. Among these two groups, the $[SeO_4]^{2-}$ do participate in the glass formation, whereas $[SeO_3]^{2-}$ act as modifying ions and generate different imperfections and cause to decrease elastic coefficients of the samples. Hence, the elastic parameters of these samples depend on relative proportions of these two groups. Quite recently, we have reported the luminescence properties of $Er^{3+}$ doped lead boroselenate glass ceramics admixed with $Au_2O_3$ [8]. $Au_2O_3$ is added with a view to enhance PL efficiency of rare earth dopant ions since the gold particles emit PL in the visible region and also bring important changes in the elastic properties of the host glass ceramic. The studies further suggested that higher the fraction of $[SeO_3]^{2-}$ units (determined by XPS studies) larger is the PL efficiency of erbium ions. Such higher efficiency was attributed to the decreased phonon losses because of higher concentration of structural defects induced by $[SeO_3]^{2-}$ units. We have added about 40 mol% of PbO to the glass system to enhance the density, refractive index (which cause to increase PL efficiency) and also to minimize the sublimity and hygroscopic nature of $SeO_2$. We have also added a minimal concentration of $B_2O_3$ (10 mol%) to this glass system in order to ensure good glass formation. Further, it may be noted that the physical characteristics of amorphous systems containing $SeO_2$ are very limitedly investigated and in most of such systems small concentration of $SeO_2$ was added as a dopant. Such studies as on the glasses or glass ceramics mixed with relatively large concentration of $SeO_2$ are very rare in spite of their high suitability for the several applications for example, in electrical and optical devices as sensors and various non-linear optical (NLO) devices etc. [9-13].

The current investigation is aimed to investigate the influence of $Au_2O_3$ on elastic characteristics of $PbO-B_2O_3-SeO_2:Er_2O_3$ glass ceramics. As mentioned earlier, addition of traces of $Au_2O_3$ to these makes them useful for the designing of miniature (nanosized) optoelectronic circuits and memory devices with large data storage capacity [14]. Owing these reasons the knowledge of elastic properties of $SeO_2$ based glass system doped with $Au_2O_3$ is highly helpful. In general, $Au_2O_3$ is a volatile oxide and breaks into AuO, $Au_2O$ and $Au^0$ metallic particles (MPs) during the synthesis of the glasses. The presence of larger content of $Au^0$ MPs will have a strong bearing on elastic properties as well as other physical characteristics of the glasses. For obtaining the glass samples with larger fraction of $Au^0$ MPs (possible by reducing trivalent gold ions), in this study the synthesized glasses were heated for a time interval of 24 h at the crystallization temperature. Such treatment reduces $Au^{3+}$ ions into $Au^0$ particles; in addition, it is quite likely for the formation of anisotropic crystalline phases e.g., $Au_2(Se^{(IV)}O_3)_3$ and $Au_2(Se^{(IV)}O_3)_2(Se^{(VI)}O_4)$ of selenium ions [15,16] in the samples. Glasses with such crystallites are useful for NLO devices [16]. Further, the development of such crystalline phases along with $Au^0$ MPs in the glasses causes to entrench several nanosized free volume defects in the material which alter the elastic properties. We propose to determine the concentration of such defects by means of positron annihilation lifetime(PAL) spectroscopy in this study [17,18].

Thus, in continuation to our earlier investigations on the impact of $Au_2O_3$ on photoluminescence features of $PbO-B_2O_3-SeO_2:Er_2O_3$ glass ceramics, the current investigation is devoted to study the elastic properties (in combination with PAL characteristics in detail) as a function of $Au_2O_3$ content and to correlate these results with those of luminescence characteristics

## 2. Experimental

The chemical composition chosen for the present is: $40PbO-10B_2O_3-(50-x)SeO_2-0.5Er_2O_3:xAu_2O_3$ (where $x$ = 0, 0.025, 0.050, 0.075 and 0.1, all in mol%) and the samples are labeled as $EA_0$, $EA_{25}$, $EA_{50}$, $EA_{75}$ and $EA_{100}$. PbO, $H_3BO_3$, $SeO_2$, $Au_2O_3$ and $Er_2O_3$ compounds were mixed thoroughly and melted in Pt crucibles in the range of temperature 1000–1050 °C (for about 40 min) in a thermally controlled furnace. The resultant molten liquid was casted in to slabs of rectangular shape and were annealed at 350 °C. For crystallization, the samples were held at 630 °C (crystallization temperature-onset, identified from DSC studies) for 24 h and then chilled to room temperature. The densities of the samples were measured by Archimedes' principle (using o-xylene as the buoyant liquid) to an accuracy of ±0.001.The other details of the preparation of the samples and characterization techniques e.g., X-ray diffraction (XRD), IR and XPS spectra etc. were reported in our earlier communication [8]. Transmission electron microscopy (TEM) images of the prepared samples were recorded on a JEOL 2200FS instrument with FE gun operated at 200 kV. XPS measurements were performed on PHI 5000 Versa Probe ULVAC instrument with a monochromatic Al Kα (hν = 1486.6 eV) as the source. Spectra were registered with respect to C 1s peak. Positron annihilation spectra were recorded on "Ortec" positron lifetime system with a $^{22}$Na isotope (100 kBq) source. Lifetime τ of the positroniums (from which the size of the free volume space defects is determined) is evaluated to a precision of 230 ps at 18 °C in the atmosphere of RH = 35%.

For measuring the longitudinal velocity in the samples, the OPBOX 2.1 measuring device with software was used. This device controls the operation of ultrasonic heads (transmitting and receiving). The frequency of the transmitted signal was 20 MHZ. The speed of the sound is determined by measuring the passage time of the ultrasonic impulse and its echo through the glass ceramic. Shear velocities were measured with a piezoelectric composite oscillator technique [19,20] using an X-cut (0.13 MHz) cylindrical quartz transducer (2.0 cm long and 0.25 cm diameter). The dimensions of the glass ceramics used for these measurements are approximately the same as those of quartz transducer.

## 3. Results and discussion

X-ray diffractogram of the glass ceramic $EA_{25}$ is presented in Fig. 1. The diffractogram exhibited two diffraction peaks at 2θ = 29.9° and 44.0° due to the reflections from $Au_2(SeO_3)_3$ crystal phase and $Au^0$ MPs, respectively [14,16,21]. The diffractograms of all the other samples have exhibited similar peaks. Out of these two peaks, the peak due to $Au^0$ MPs exhibited a gradual growth up to 0.075 mol% of $Au_2O_3$ (inset of Fig. 1); such growth of the peak suggests an increase in the concentration of $Au^0$ metallic particles. It may be noted here that in the diffractogram of the as-prepared sample no peaks were noticed and indicated the amorphous character of such samples.

In order to prove, further, the crystallinity in the samples, we have recorded TEM images and presented in Fig.2, for the samples $EA_{50}$ and $EA_{75}$ with different magnifications. The pictures indicated a large number segregated crystal grains of nanosize embedded in the glass samples.

Further, we have recorded positron annihilation decay profiles in these glass ceramics in order to estimate the free volume defects/free volume space entrenched in the glass ceramics. Fig. 3(a) represents the positron decay profiles of one of the glass ceramic samples viz., $EA_{100}$. The decay curve is deconvoluted in to $I_1$, $I_2$ and $I_3$ components. By using the lifetime ($τ_3$) of the third component (long lived component), we have estimated radius R of free cavities in the glass ceramics with the standard expression [22,23]:

$$\tau_3 = \frac{1}{2}\left[1 - \frac{R}{R_0} + \frac{1}{2\pi}\sin\left(\frac{2\pi R}{R_0}\right)\right]^{-1} \quad (1)$$

Using the value of $I_3$ (the intensity of the 3$^{rd}$ component of the decay profiles) and the radius R of free cavities (found to be in the range of 2.6 to 2.8 Å), the free volume space fraction $f_v$ trapped in the sample is determined with the equation:

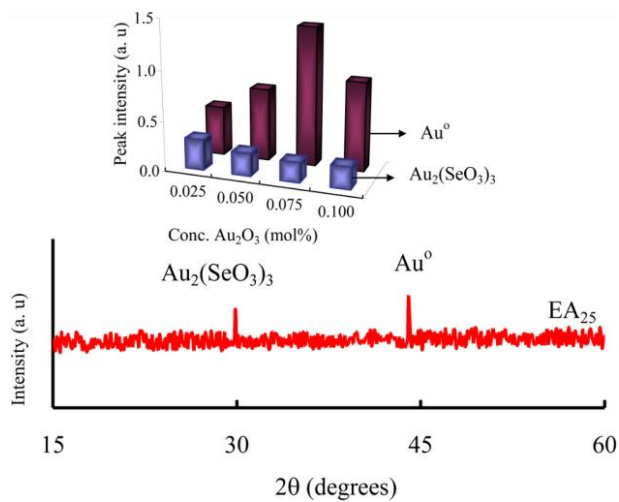

**Fig. 1.** X-ray diffractogram of $EA_{25}$ sample [Ref. 8]. Inset represents the variation of the intensity of the diffraction peak related to $Au^0$ metallic particles and $Au_2(SeO_3)_3$ crystalline phase with the concentration of $Au_2O_3$. The patterns were recorded in continuous scan mode with a scan speed of 2 deg/min.

A. Siva Sesha Reddy et al.

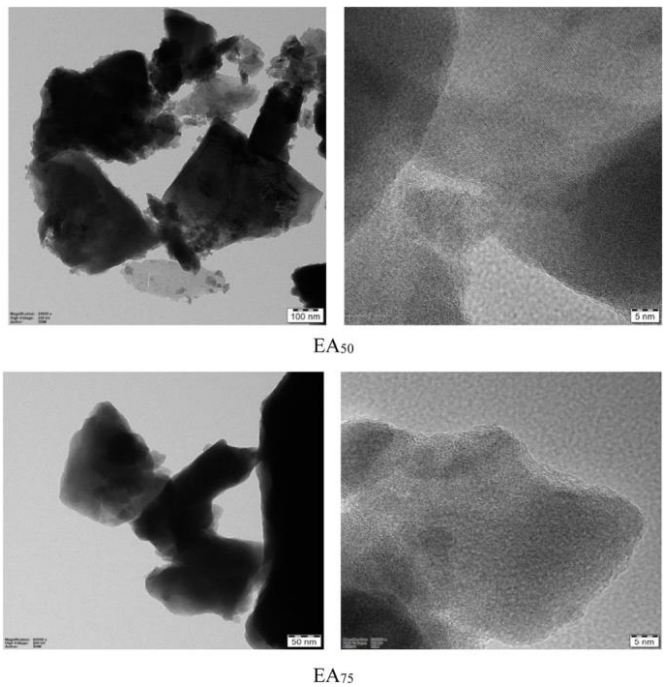

**Fig. 2.** TEM images of EA$_{50}$ and EA$_{75}$ glass ceramic samples recorded with different magnifications.

$$f_v = AI_3(\%)\left(\frac{4}{3}\pi R^3\right) \quad (2)$$

In Eq. (2), relative fraction of the f$_v$ in the titled samples (without using the value of A whose value is predicted for polymers as 0.0018/Å$^3$) is estimated with respect to that of the pure glass, using following equation

$$\frac{\Delta f}{f} = \left(\frac{f_{v_i} - f_{v_0}}{f_{v_0}}\right) \cdot 100\% \quad (3)$$

In Eq. (3), f$_{vi}$ represents free volume of the titled glass ceramics mixed with different contents of gold oxide and f$_{v0}$ denotes free volume in the pure glass ceramic.

The free volume fraction of the free space ($\Delta f/f_0$) obtained relative to that of the un-doped samples is plotted against Au$_2$O$_3$ concentration in Fig. 3(b). The plot showed a gradual increase of the free space fraction in the samples up to 0.075 mol% of Au$_2$O$_3$. Such increase of magnitude of free space obviously causes to enhance the acoustic impedance of the glass ceramics leading to decrease the values of elastic parameters.

In order to have additional confirmation of the presence of Au$^0$ MPs (which act as modifiers) and to have an idea over their variation with the concentration of Au$_2$O$_3$ in these samples we have recorded XPS spectra. Fig. 4 represents such spectrum for the sample EA$_{100}$. The spectrum exhibited the peaks corresponding to the binding energy (BE) of 4f Au$^{3+}$ ions at 85.8 eV and 89.6 eV and two more peaks at 84.0 eV and 87.6 eV connected with the BE of Au$^0$ MPs [24,25]. Fig.4 (inset) represents variation of concentration of Au$^{3+}$ ions and Au$^0$ MPs with the content of Au$_2$O$_3$. The figure showed an increase of Au$^0$ MPs (at the expense of Au$^{3+}$ions) up to 0.075 mol% of Au$_2$O$_3$. These gold MPs behave as modifiers and produce structural imperfections as mentioned earlier [8,17].

Fig.5, represents the infrared (IR) spectrum of PbO–B$_2$O$_3$–SeO$_2$–Er$_2$O$_3$ glass ceramic containing 0.05 mol% of Au$_2$O$_3$. Spectrum exhibited the vibrational bands due to [SeO$_4$]$^{2-}$ units (at nearly 950 cm$^{-1}$) and stretching (symmetrical) vibrational band of quarantined selenite [SeO$_3$]$^{2-}$ pyramidal groups in the wavenumber range 850-900 cm$^{-1}$ [11]. In addition, the spectrum also exhibited the conventional vibrational bands of BO$_3$ and BO$_4$ units as shown in the figure. With increase of Au$_2$O$_3$ content up to 0.075 mol%, the vibrational band of [SeO$_3$]$^{2-}$ groups exhibited a gradual growth, whereas that of [SeO$_4$]$^{2-}$ groups showed a decaying trend (inset of Fig. 5). The selenite [SeO$_3$]$^{2-}$ groups mainly act as modifiers [12] and induce structural defects in the glass ceramic. In support of this argument, we have observed the growth of BO$_3$ vibrational band, whereas that of BO$_4$ units exhibited a decreasing trend

Fig. 6(a) and (b) represent propagation of longitudinal waves (by which the velocity v$_l$ is determined) in the samples EA$_0$ and EA$_{50}$. In these figures, measuring gates are represented by the two arrows. Such arrows in fact, indicate the starting and end point of the measuremen between two consecutive reflected pulse

Using the values of v$_l$ and v$_s$, various elastic coefficients of PBSE glass ceramics doped with different concentrations of Au$_2$O$_3$ are estimated with the standard expressions and tabulated in Table 1. Their variations with the concentration of Au$_2$O$_3$ are presented in Fig. 7.

All these parameters showed a decreasing trend with the content of Au$_2$O$_3$ up to 0.075 mol%. In Fig. 8(a) and (b), Young's and shear moduli variations with the content of Au$_2$O$_3$ are presented. The variations showed the gradual decrement of these parameters (and also the microhardness H of the samples (Table 1)) with the content of Au$_2$O$_3$.

Zhang et al. [26] have established a general empirical relation between Vickers hardness H$_v$ and yield strength $\sigma_y$ as

$$H_y \approx 3\sigma_y \quad (4)$$

The obtained vales of $\sigma_y$ evaluated based on this relation are presented in Table 1. The value of $σ_y$ exhibited a decreasing trend with the concentration of Au$_2$O$_3$ up to 0.075 mol% similar to Vickers hardness H$_v$ and subsequently, Alao and Yin [27] have applied this relation to evaluate the dependence of contact hardness, H$_c$, on the yield strength, $σ_y$, for lithium metasilicate glass–ceramics. According to these people the value of H$_c$ is given by

$$H_c = Cσ_y \quad (5)$$

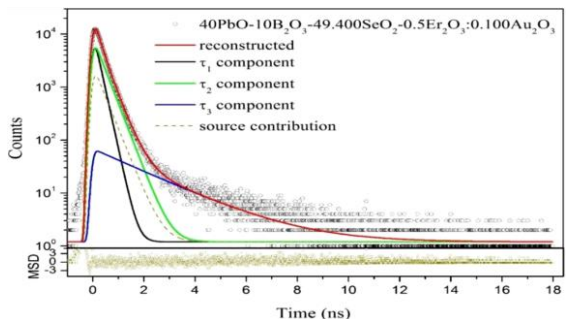

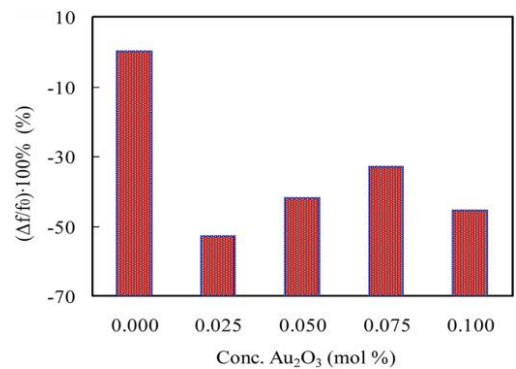

**Fig. 3.** (a) Positron annihilation lifetime decay profiles of the glass ceramics EA$_{100}$ doped PBSE glass ceramics. Fig. 3(b) Variations of free volume fraction $\Delta f/f_0$ with concentrations of Au$_2$O$_3$

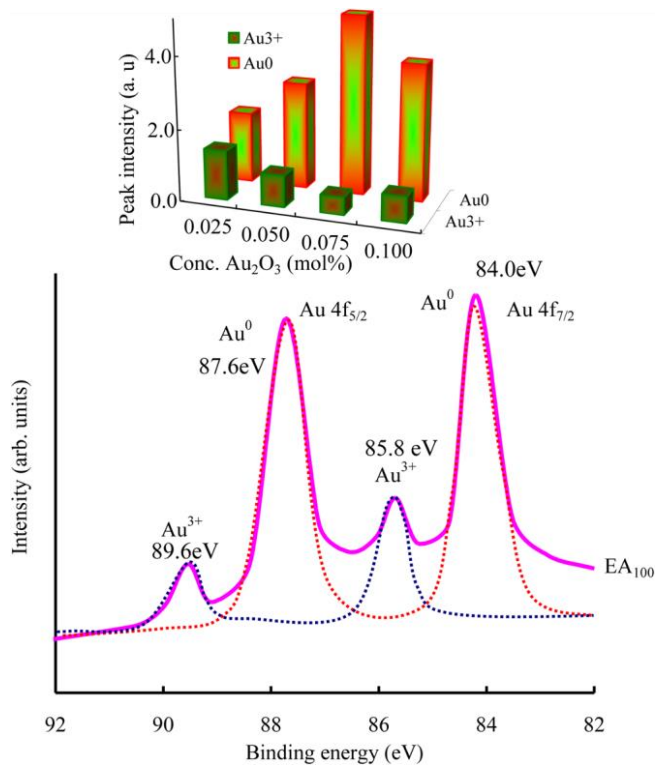

**Fig. 4.** XPS spectrum of EA$_{100}$ sample. Inset shows the relative variations of intensities of the peaks of Au$^0$ and Au$^{3+}$ ions with content of Au$_2$O$_3$.

In Eq. (5), C is a constraint factor which is found to be 3 for materials that exhibit a combination of piling-up and shear-banding phenomenon such as lithium metasilicate glass–ceramics [26]. In the present case for PBSE glass ceramics the evaluated value of H$_v$ gives an approximate idea over the dependence of yield strength $σ_y$ on the concentration of Au$_2$O$_3$. In other words it can be said that the yield strength is minimal for the glass ceramic EA$_{75}$.

The IR spectra have clearly suggested that an increasing concentration of [SeO$_3$]$^{2-}$ structural groups with increase of Au$_2$O$_3$. XPS spectra have demonstrated a steady growth in the concentration of Au$^0$ metallic particles.

Both of these acted as modifiers and de-augment Se-O-Se linkages in the glass network and also introduce free volume defects in the glass ceramics. Such enhanced degree of de-fragmentation in the glass network facilitated a decrease of ultrasonic speed and caused to decrease elastic parameters and micro-hardness.

Our earlier results on photoluminescence studies of these glass ceramics have also indicated an increase in the luminescence efficiency of green emission ($^4S_{3/2} → \, ^4I_{15/2}$) and orange emission ($^4F_{9/2} → \, ^4I_{15/2}$) of erbium ions, as mentioned earlier, with increase of Au$_2$O$_3$ content [8] and patterns of PL emission and band positions obtained were found to be in accordance with the reported data in the literature [28–30]. Such enhancement was attributed to the decrease of phonon losses or due to the rise in the degree of internal structural chaos in the glass ceramic network. In conclusion even though, the introduction of Au$_2$O$_3$ is proved to improve the optical properties and also the electrical characteristics in various glasses and glass ceramics, it caused to decrease the elastic coefficients and micro-hardness of PbO-B$_2$O$_3$-SeO$_2$:Er$_2$O$_3$ glass ceramics.

### 4. Conclusions

Different elastic coefficients and micro-hardness H/yield strength $σ_y$ of Au$_2$O$_3$ mixed PbO–B$_2$O$_3$–SeO$_2$:Er$_2$O$_3$ glass ceramics were estimated using ultrasonic velocities. The coefficients were found to decrease with Au$_2$O$_3$ concentration up to 0.075 mol%. Such decrease was ascribed to growing presence of [SeO$_3$]$^{2-}$ structural groups and also Au$^0$ metallic particles that produce structural imperfections like free volume defects by acting as modifiers. This argument was further supported by XPS, IR and positron annihilation spectroscopy studies. Such increased degree of disorder or increased magnitude of free volume space with increase of Au$_2$O$_3$ content was projected to be an obstruction for the passage of ultrasonic waves in the samples and hence responsible for the decrement in the elastic coefficients and micro-hardness of the studied glass ceramics. Finally, it is concluded that even though, the Au$_2$O$_3$ is a favorable dopant for superior optical and electrical properties of the studied glass ceramics it caused to have an adverse effect on the elastic coefficients. However, when the concentration of Au$_2$O$_3$ is increased beyond 0.075 mol%, we have observed a slight increase of elastic coefficients and micro-hardness.

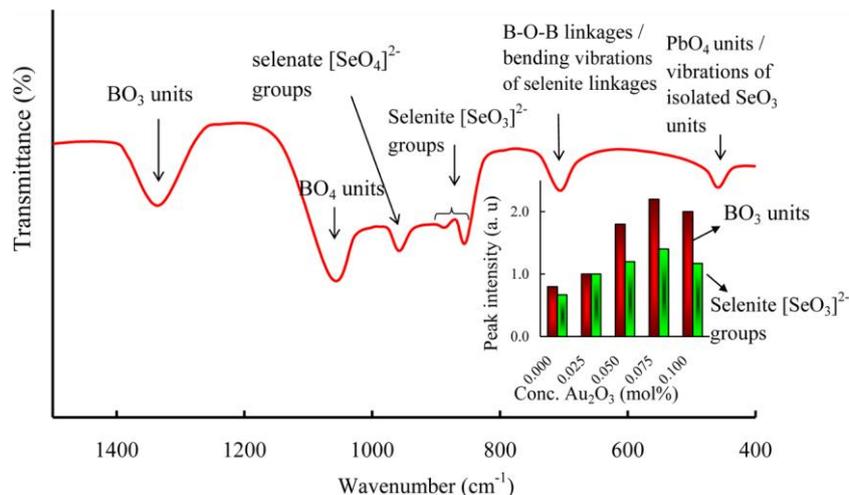

**Fig. 5.** IR spectrum EA$_{50}$ glass ceramic. In the inset relative variations of intensities of the various vibrational groups are presented with the content of Au$_2$O$_3$.

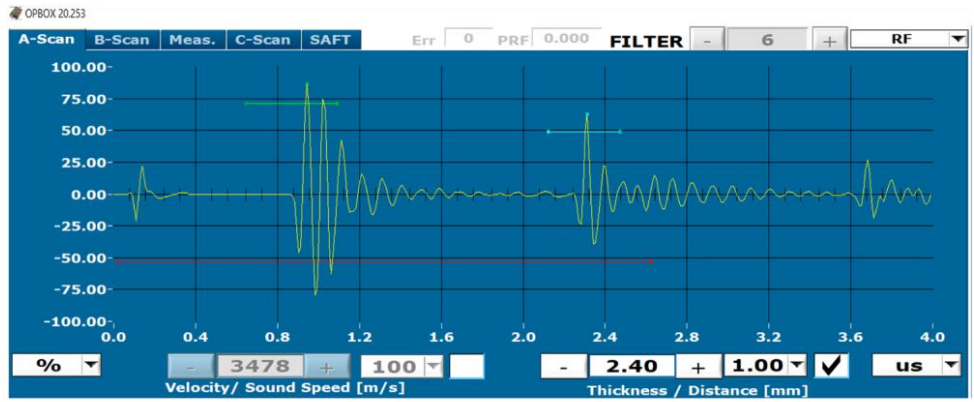

EA$_0$

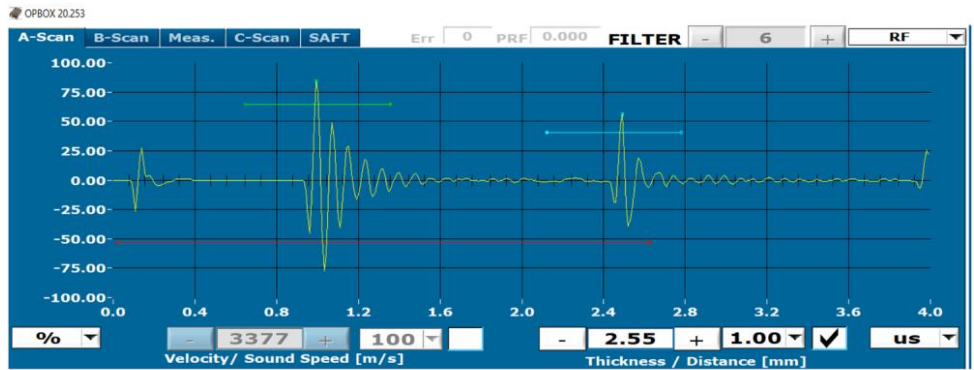

EA$_{50}$

**Fig. 6.** Longitudinal wave propagation in the samples (a) EA$_0$ and (b) EA$_{50}$.

**Table 1**
Elastic and acoustic properties of Au$_2$O$_3$ doped PBSE glass ceramics.

| Sample | Long. Vel. $v_l$ (m/s) | Density $\rho$ (g/cm$^3$) | Long. coeff. (x10$^{10}$, N/m$^2$) $L = \rho v_l^2$ | Shear Vel. $v_s$ (m/s) | Shear Modulus, G (x10$^{10}$, N/m$^2$) $G = \rho v_s^2$ | Young's Modulus, Y (x10$^{10}$ N/m$^2$) $Y = (1+\sigma)2G$ | Avg. Vel. $V_m$ (m/s) | Debye Temp. $\theta_D$ (K) | Micro hardness H (x10$^9$, N/m$^2$) $H = \frac{(1-2\sigma)Y}{6(1+\sigma)}$ | Yield strength $\sigma_y$ (GPa) |
|---|---|---|---|---|---|---|---|---|---|---|
| EA$_0$ | 3478 | 5.87 | 7.10 | 2333 | 3.19 | 7.00 | 2200 | 99.8 | 8.71 | 2.90 |
| EA$_{25}$ | 3404 | 5.92 | 6.86 | 2283 | 3.09 | 6.73 | 2153 | 97.9 | 8.42 | 2.81 |
| EA$_{50}$ | 3377 | 5.94 | 6.77 | 2265 | 3.05 | 6.65 | 2136 | 97.3 | 8.31 | 2.77 |
| EA$_{75}$ | 3333 | 6.09 | 6.77 | 2235 | 3.04 | 6.64 | 2108 | 96.8 | 8.30 | 2.76 |
| EA$_{100}$ | 3375 | 6.10 | 6.95 | 2264 | 3.13 | 6.72 | 2134 | 98.1 | 8.53 | 2.84 |

*$\sigma$ is the Poison's ratio, $V_m = v_m = [v_l^2 - (4/3)v_s^2]^{1/2}$ and $\theta_D = \theta_D = \frac{h}{k_B}\left(\frac{3N_A}{4\pi V_s}\right)^{1/3} v_m$

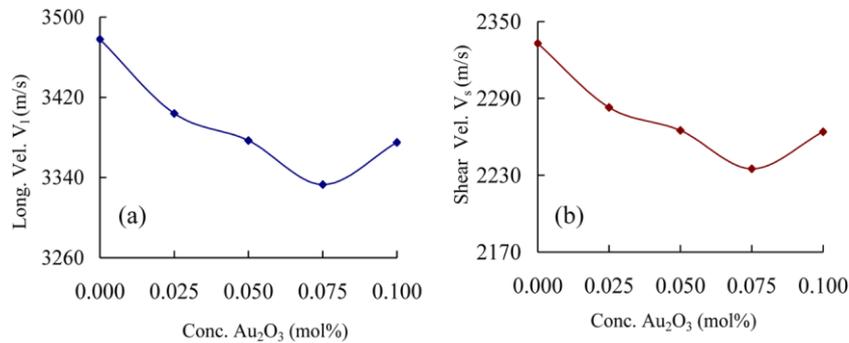

**Fig. 7.** Variations of (a) $V_l$ and (b) $V_s$ with the concentration of Au$_2$O$_3$ in the titled glass ceramics.

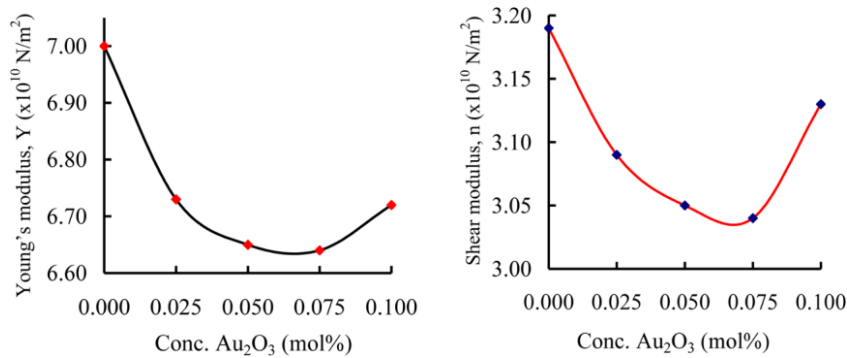

**Fig. 8.** The plots of elastic coefficients vs concentration of $Au_2O_3$.

### CRediT authorship contribution statement

**A. Siva Sesha Reddy:** Conceptualization, Methodology, Investigation. **A.V. Kityk:** Methodology, Data curation, Formal analysis, Software, Writing – original draft. **J. Jedryka:** Methodology, Data curation, Formal analysis, Software, Writing – original draft. **P. Rakus:** Methodology, Data curation, Formal analysis, Software, Writing – original draft. **A. Wojciechowski:** Methodology, Data curation, Formal analysis, Software, Writing – original draft. **N. Venkatramaiah:** Methodology, Data curation, Formal analysis. **V. Ravi Kumar:** Conceptualization, Methodology, Investigation, Supervision, Writing – original draft, Writing – review & editing. **N. Veeraiah:** Supervision, Writing – original draft, Writing – review & editing.

### Acknowledgments

Some of the presented results are part of a project that has received funding from the European Union's Horizon 2020 research and innovation programme under the Marie Skodowska-Curie grant agreement No 778156. A.V.K., J.J., P.R and A.W. acknowledge support from resources for science in years 2018-2022 granted for the realization of international co-financed project Nr W13/H2020/2018 (Dec. MNiSW 3871/H2020/2018/2).

### References

[1] Y. Liu, D. Yang, L. Riekehr, H. Engqvist, L. Fud, W. Xia, Combining good mechanical properties and high translucency in yttrium-doped ZrO2-SiO2 nanocrystalline glass-ceramics, J. Eur. Ceram. Soc. 42 (2022) 274–285.

[2] Ch.Gautam Shweta, V.P. Tripathi, S. Kumar, S. Behera, R.K. Gautam, Synthesis, physical and mechanical properties of lead strontium titanate glass ceramics, Phy. B 615 (2021), 413069.

[3] A. Venkata Sekhar, A. Siva Sesha Reddy, A. Kityk, J. Jedryka, P. Rakus, A. Wojciechowski, G. Naga Raju, N. Veeraiah, Influence of NiO doping on elastic properties of Li2SO4-MgO-P2O5 glass system- Investigation by means of acoustic wave propagation, Appl. Phys. A 127 (2021) 342.

[4] Abd El-Moneim, Md. Eltohamy, H. Afifi, M.S. Gaafar, A. Atef, An ultrasonic study on ternary xPbO–(45-x)CuO–55B2O3 glasses, Ceram. Int. 47 (2021) 27351–27360.

[5] N. Narasimha Rao, I.V. Kityk, V. Ravi Kumar, P. Raghava Rao, B.V. Raghavaiah, P. Czaja, P. Rakus, N. Veeraiah, Piezoelectric and elastic properties of ZnF2–PbO–TeO2: tiO2 glass ceramics, J. Non Cryst. Solids 358 (2012) 702–710.

[6] S. Thakur, A. Kaur, L. Singh, Mixed valence effect of Se6+ and Zr4+ on structural, thermal, physical, and optical properties of B2O3–Bi2O3–SeO2–ZrO2 glasses, Opt. Mater. 96 (2019), 109338.

[7] Q. Chen, K. Su, Z. Zhao, Q. Ma, Optical and electrical properties of SeO2 modified PbO-Bi2O3-B2O3 glasses, J. Non Cryst. Solids 498 (2018) 448–454.

[8] A. Siva Sesha Reddy, Valluri Ravi Kumar, G. Lakshminarayana, N. Purnachand, N. Venkatramaiah, V. Ravi Kumar, N. Veeraiah, Influence of gold ions on visible and NIR luminescence features of Er3+ ions in lead boroselenate glass ceramics, J. Lumin. 226 (2020), 117481.

[9] A. Palui, A. Ghosh, Structure and dielectric properties of Ag2O-SeO2-TeO2 mixed former glasses, J. Non Cryst. Solids 482 (2018) 230–235.

[10] A. Palui, A. Ghosh, Mixed glass former effect in Ag2O SeO2 TeO2 glasses: dependence on characteristic displacement of mobile ions and relative population of bond vibrations, J. Phys. Chem. C 121 (2017) 8738–8745.

[11] L. Lakov, S. Yordanov, Preparation and study of new class SeO2 based glass obtained at high oxygen pressure, Mach. Technol. Mater. 15 (2021) 38–41.

[12] Q. Chen, K. Su, Z. Zhao, Q. Ma, Optical and electrical properties of SeO2 modified PbO-Bi2O3-B2O3 glasses, J. Non Cryst. Solids 498 (2018) 448–454.

[13] Q. Chen, K. Su, Y. Li, Z. Zhao, Structure, spectra and thermal, mechanical, faraday rotation properties of novel diamagnetic SeO2-PbO-Bi2O3-B2O3 glasses, Opt. Mater. 80 (2018) 216–224.

[14] R. Rajaramakrishna, S. Karuthedath, R.V. Anavekar, H. Jain, Nonlinear optical studies of lead lanthanum borate glass doped with Au nanoparticles, J. Non Cryst. Solids 358 (2012) 1667–1672.

[15] R.H. Lambertson, C.A. Lacy, S.D. Gillespie, M.C. Leopold, R.H. Coppage, Gold nanoparticle colorants as traditional ceramic glaze alternatives, J. Am. Ceram. Soc. 100 (2017) 3943–3951.

[16] K. Virender Sharma, T.J. Mc-Donald, M. Sohn, G.A.K. Anquandah, M. Pettine, R. Zboril, Biogeochemistry of selenium. A review, Environ. Chem. Lett. 13 (2014) 49–58.

[17] J. Ashok, M. Kostrzewa, A. Ingram, M. Srinivasa Reddy, V. Ravi Kumar, Y. Gandhi, N. Veeraiah, Free volume estimation in Au and Ag mixed sodium antimonate glass ceramics by means of positron annihilation, Phys. B 570 (2019) 266–273.

[18] A. Siva Sesha Reddy, M. Kostrzewa, A. Ingram, N. Purnachand, P. Bragiel, V. Ravi Kumar, I.V. Kityk, N. Veeraiah, Positron annihilation exploration of voids in zinc zirconium borate glass ceramics entrenched with ZnZrO3 perovskite crystal phases, J. Eur. Ceram. Soc. 38 (2018) 2010–2016.

[19] A.V. Ravi Kumar, N. Veeraiah, Acoustic Investigations on LiF-B2O3-glasses doped with certain rare earth ions, J. Mater. Sci. Lett. 18 (1999) 475–478.

[20] M. Rami Reddy, S.B. Raju, N. Veeraiah, Acoustic investigations on PbO-Al2O3-B2O3 glasses doped with certain rare-earth ions, Bull. Mater. Sci. 24 (2001) 63–68.

[21] P.G. Jones, G.M. Sheldrick, E. Schwarzmann, A. Vielmäder, Preparation and crystal structure of di-gold(III) bis(selenite)oxide, Au2(SeO3)2O, Z. Naturforsch. 38 (1983) 10–11.

[22] S.J. Tao, Positronium annihilation in molecular substances, J. Chem. Phys. 56 (1972) 5499.

[23] M. Eldrup, D. Lighbody, J.N. Sherwood, The temperature dependence of positron lifetimes in solid pivalic acid, Chem. Phys. 63 (1981) 51–58.

[24] Y. Mikhlin, M. Likhatski, Y. Tomashevich, A. Romanchenko, S. Erenburg, S. Trubina, XAS and XPS examination of the Au-S nanostructures produced via the reduction of aqueous gold (III) by sulfide ions, J. Electron. Spectrosc. Relat. Phenom. 177 (2010) 24–29.

[25] M. Zacharska, A.L. Chuvilin, V.V. Kriventsov, S. Beloshapkin, M. Estrada, A. Simakov, D.A. Bulushev, Support effect for nanosized Au catalysts in hydrogen production from formic acid decomposition, Catal. Sci. Technol. 6 (2016) 6853–6860.

[26] P. Zhang, S.X. Li, Z.F. Zhang, General relationship between strength and hardness, Mater. Sci. Eng. A 529 (2011) 62–73.

[27] A.R. Alao, L. Yin, Nano-mechanical behaviour of lithium metasilicate glass–ceramic, J. Mech. Behav. Biomed. Mater. 49 (2015) 162–174.

[28] C.R. Kesavulu, V.B. Sreedhar, C.K. Jayasankar, et al., Structural, thermal and spectroscopic properties of highly Er3+ doped novel oxyfluoride glasses for photonic application, Mater. Res. Bull. 51 (2014) 336–344.

[29] Ch. Basavapoornima, K. Linganna, C.R. Kesavulu, S. Ju, B.H. Kim, W.T. Han, C. K. Jayasankar, Spectroscopic and pump power dependent upconversion studies of Er3+ doped lead phosphate glasses for photonic applications, J. Alloys Compd. 699 (2017) 959–968.

[30] F. Rivera-López, P. Babu, L. Jyothi, U.R. Rodríguez-Mendoza, I.R. Martín, C. K. Jayasankar, V. Lavín, Er3+–Yb3+ codoped phosphate glasses used for an efficient 1.5 μm broadband gain medium, Opt. Mater. 34 (2012) 1235–1240.